# Suppressing Random Walks in Markov Chain Monte Carlo Using Ordered Overrelaxation

Radford M. Neal

Dept. of Statistics and Dept. of Computer Science
University of Toronto
Toronto, Ontario, Canada

World Wide Web: http://www.cs.toronto.edu/~radford
E-mail: radford@stat.toronto.edu

21 June 1995

Markov chain Monte Carlo methods such as Gibbs sampling and simple forms of the Metropolis algorithm typically move about the distribution being sampled via a random walk. For the complex, high-dimensional distributions commonly encountered in Bayesian inference and statistical physics, the distance moved in each iteration of these algorithms will usually be small, because it is difficult or impossible to transform the problem to eliminate dependencies between variables. The inefficiency inherent in taking such small steps is greatly exacerbated when the algorithm operates via a random walk, as in such a case moving to a point $n$ steps away will typically take around $n^2$ iterations. Such random walks can sometimes be suppressed using "overrelaxed" variants of Gibbs sampling (a.k.a. the heatbath algorithm), but such methods have hitherto been largely restricted to problems where all the full conditional distributions are Gaussian. I present an overrelaxed Markov chain Monte Carlo algorithm based on order statistics that is more widely applicable. In particular, the algorithm can be applied whenever the full conditional distributions are such that their cumulative distribution functions and inverse cumulative distribution functions can be efficiently computed. The method is demonstrated on an inference problem for a simple hierarchical Bayesian model.

# 1  Introduction

Markov chain Monte Carlo methods are used to estimate the expectations of various functions of a state, $x = (x_1, \ldots, x_N)$, with respect to a distribution given by some density function, $\pi(x)$. Typically, the dimensionality, $N$, is large, and the density $\pi(x)$ is of a complex form, in which the components of $x$ are highly dependent. The estimates are based on a (dependent) sample of states obtained by simulating an ergodic Markov chain that has $\pi(x)$ as its equilibrium distribution. Starting with the work of Metropolis, et al. (1953), Markov chain Monte Carlo methods have been widely used to solve problems in statistical physics and, more recently, Bayesian statistical inference. It is often the only approach known that is computationally feasible. Various Markov chain Monte Carlo methods and their applications are reviewed by Toussaint (1989), Neal (1993), and Smith and Roberts (1993).

For the difficult problems that are their primary domain, Markov chain Monte Carlo methods are limited in their efficiency by strong dependencies between components of the state, which force the Markov chain to move about the distribution in small steps. In the widely-used Gibbs sampling method (known as the heatbath method in the physics literature), the Markov chain operates by successively replacing each component of the state, $x_i$, by a value randomly chosen from its conditional distribution given the current values of the other components, $\pi(x_i \mid \{x_j\}_{j \neq i})$. When dependencies between variables are strong, these conditional distributions will be much narrower than the corresponding marginal distributions, $\pi(x_i)$, and many iterations of the Markov chain will be necessary for the state to visit the full range of the distribution defined by $\pi(x)$. Similar behaviour is typical when the Metropolis algorithm is used to update each component of the state in turn, and also when the Metropolis algorithm is used with a simple proposal distribution that changes all components of the state simultaneously.

This inefficiency due to dependencies between components is to a certain extent unavoidable. We might hope to eliminate the problem by transforming to a parameterization in which the components of the state are no longer dependent. If this can easily be done, it is certainly the preferred solution. Typically, however, finding and applying such a transformation is difficult or impossible. Even for a distribution as simple as a multivariate Gaussian, eliminating dependencies will not be easy if the state has millions of components, as it might for a problem in statistical physics or image processing.

However, in the Markov chain Monte Carlo methods that are most commonly used, this inherent inefficiency is greatly exacerbated by the random walk nature of the algorithm. Not only is the distribution explored by taking small steps, the direction of these steps is randomized in each iteration, with the result that on average it takes about $n^2$ steps to move to a point $n$ steps away. This can greatly increase both the number of iterations required before equilibrium is approached, and the number of subsequent iterations that are needed to gather a sample of states from which accurate estimates for the quantities of interest can be obtained.

In the physics literature, this problem has been addressed in two ways — by "overrelaxation" methods, introduced by Adler (1981), which are the main subject of this paper, and by dynamical methods, such as "hybrid Monte Carlo", which I briefly describe next.

The hybrid Monte Carlo method, due to Duane, Kennedy, Pendleton, and Roweth (1987), can be seen as an elaborate form of the Metropolis algorithm (in an extended state space) in which candidate states are found by simulating a trajectory defined by Hamiltonian dynamics. These trajectories will proceed in a consistent direction, until such time as they reach a region of low probability. By using states proposed by this deterministic process, random walk effects can be largely eliminated. In Bayesian inference problems for complex models based on neural networks, I have found (Neal 1995) that the hybrid Monte Carlo method can be hundreds or thousands of times faster than simple versions of the Metropolis algorithm.

Hybrid Monte Carlo can be applied to a wide variety of problems where the state variables are continuous, and derivatives of the probability density can be efficiently computed. The method does, however, require that careful choices be made both for the length of the trajectories and for the stepsize used in the discretization of the dynamics. Using too large a stepsize will cause the dynamics to become unstable, resulting in an extremely high rejection rate. This need to carefully select the stepsize in the hybrid Monte Carlo method is similar to the need to carefully select the width of the proposal distribution in simple forms of the Metropolis algorithm. (For example, if a candidate state is drawn from a Gaussian distribution centred at the current state, one must somehow decide what the standard deviation of this distribution should be). Gibbs sampling does not require that the user set such parameters. A Markov chain Monte Carlo method that shared this advantage while also suppressing random walk behaviour would therefore be of interest.

Markov chain methods based on "overrelaxation" show promise in this regard. The original overrelaxation method of Adler (1981) is similar to Gibbs sampling, except that the new value chosen for a component of the state is negatively correlated with the old value. In many circumstances, successive overrelaxation improves sampling efficiency by suppressing random walk behaviour. Like Gibbs sampling, Adler's overrelaxation method does not require that the user select a suitable value for a stepsize parameter. It is therefore significantly easier to use than hybrid Monte Carlo (although one does still need to set a parameter that plays a role analogous to the trajectory length in hybrid Monte Carlo). Overrelaxation methods also do not suffer from the growth in computation time with system size that results from the use of a global acceptance test in hybrid Monte Carlo. (On the other hand, although overrelaxation has been found to greatly improve sampling in a number of problems, there are distributions for which overrelaxation is ineffective, but hybrid Monte Carlo works well.)

Unfortunately, Adler's original overrelaxation method is applicable only to problems where all the full conditional distributions are Gaussian. Several proposals have been made for overrelaxation methods that are more generally applicable (see Section 3 below). Most of these methods employ occasional rejections to ensure that the correct distribution is invariant. As we will see, however, such rejections can undermine the ability of overrelaxation to suppress random walks. Moreover, the probability of rejection in these methods is determined by the distribution to be sampled, and cannot be reduced in any obvious way.

In this paper, I present a rejection-free overrelaxation method based on the use of order statistics. In principle, this "ordered overrelaxation" method can be used to sample from any distribution for which Gibbs sampling would produce an ergodic Markov chain. In practice, the method will be useful only if the required computations can be performed efficiently. I discuss in detail one strategy for performing these computations, which is applicable to problems where the full conditional distributions have forms for which the cumulative distribution functions and inverse cumulative distribution functions can be efficiently computed. I also mention several other strategies that may further widen the range of problems to which ordered overrelaxation can be applied.

In Section 2, which follows, I review Adler's Gaussian overrelaxation method. In Section 3, I discuss previous proposals for overrelaxation methods that are more generally applicable. The new method of ordered overrelaxation is introduced in Section 4, and strategies for its implementation are discussed in Section 5. In Section 6, the strategy employing cumulative distribution functions is used to demonstrate the ordered overrelaxation method on a Bayesian inference problem for a simple hierarchical model. I conclude by discussing how the method might be applied in practice, and some possibilities for future work.

## 2  Overrelaxation with Gaussian conditional distributions

Overrelaxation methods have long been used in the iterative solution of systems of linear equations (Young 1971), and hence also for the minimization of quadratic functions. The first Markov chain sampling method based on overrelaxation was introduced in the physics literature by Adler (1981), and later studied by Whitmer (1984). The same method was later found by Barone and Frigessi (1989), and discussed in a statistical context by Green and Han (1992). Though itself limited to problems with Gaussian conditional distributions, Adler's method is the starting point for the more general methods that have since been proposed.

### 2.1  Adler's Gaussian overrelaxation method

Adler's overrelaxation method is applicable when the distribution for the state, $x = (x_1, \ldots, x_N)$, is such that the full conditional densities, $\pi(x_i \,|\, \{x_j\}_{j \neq i})$, are all Gaussian (in the terminology used by Adler, when the log probability density is "multiquadratic"). Note that this class includes distributions other than the multivariate Gaussians, such as $\pi(x_1, x_2) \propto \exp(-(1 + x_1^2)(1 + x_2^2))$. As in Gibbs sampling, the components of the state are updated in turn, using some fixed ordering. The new value chosen for component $i$ will depend on its conditional mean, $\mu_i$, and variance, $\sigma_i^2$, which in general are functions of the other components, $x_j$ for $j \neq i$. In Adler's method, the old value, $x_i$, is replaced by the new value

$$x_i' \;=\; \mu_i + \alpha\,(x_i - \mu_i) + \sigma_i\,(1 - \alpha^2)^{1/2}\,n \tag{1}$$

where $n$ is a Gaussian random variate with mean zero and variance one. The parameter $\alpha$ controls the degree of overrelaxation (or underrelaxation); for the method to be valid, we must have $-1 \leq \alpha \leq +1$. Overrelaxation to the other side of the mean occurs when $\alpha$ is negative. When $\alpha$ is zero, the method is equivalent to Gibbs sampling. (In the literature, the method is often parameterized in terms of $\omega = 1 - \alpha$. I have not followed

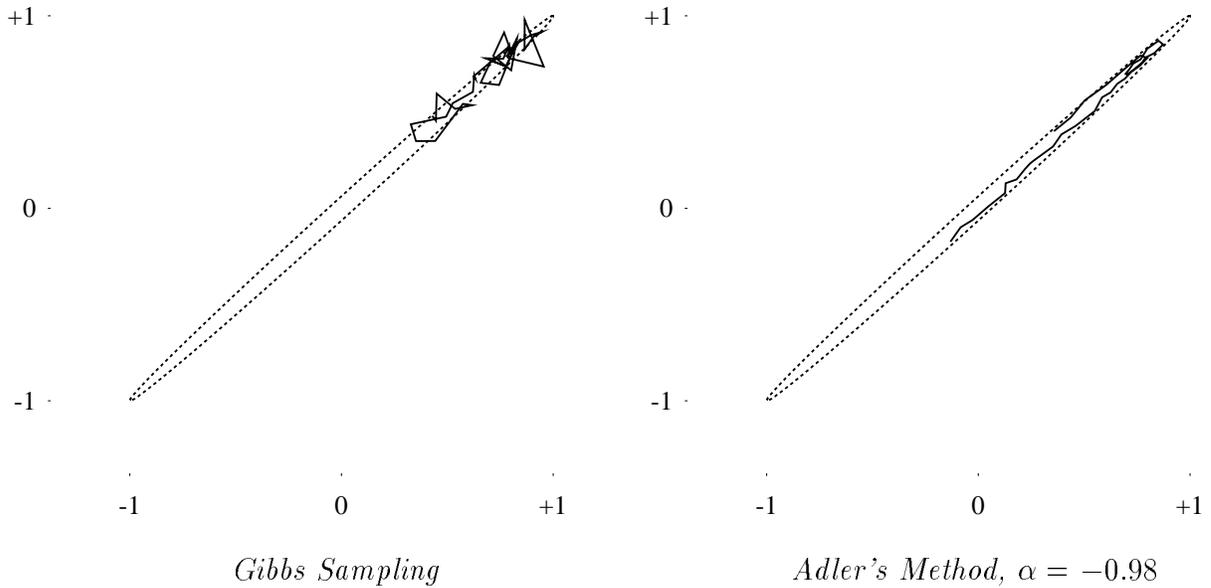

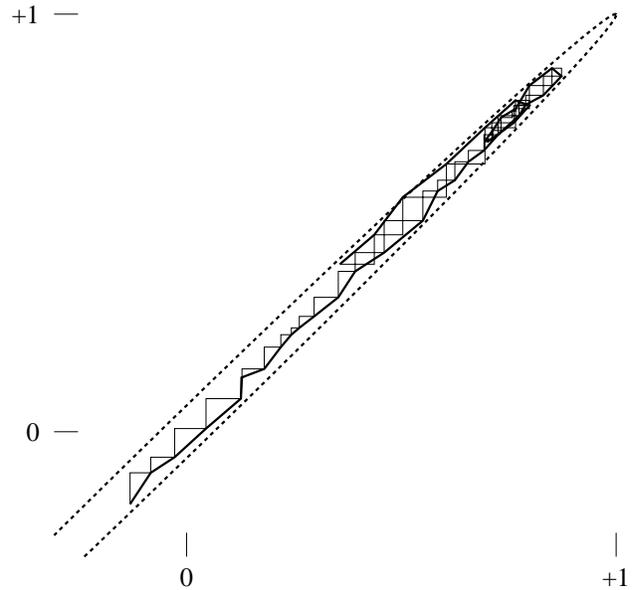

Figure 1: Gibbs sampling and Adler's overrelaxation method applied to a bivariate Gaussian with correlation 0.998 (whose one-standard-deviation contour is plotted). The top left shows the progress of 40 Gibbs sampling iterations (each consisting of one update for each variable). The top right shows 40 overrelaxed iterations, with $\alpha = -0.98$. The close-up on the right shows how successive overrelaxed updates operate to avoid a random walk.

this convention, as it appears to me to make all the equations harder to understand.)

One can easily confirm that Adler's method leaves the desired distribution invariant — that is, if $x_i$ has the desired distribution (Gaussian with mean $\mu_i$ and variance $\sigma_i^2$), then $x_i'$ also has this distribution. Furthermore, it is clear that overrelaxed updates with $-1 < \alpha < +1$ produce an ergodic chain. When $\alpha = -1$ the method is not ergodic, though updates with $\alpha = -1$ can form part of an ergodic scheme in which other updates are performed as well, as in the "hybrid overrelaxation" method discussed by Wolff (1992).

## 2.2 How overrelaxation can suppress random walks

The effect of overrelaxation is illustrated in Figure 1, in which both Gibbs sampling and the overrelaxation method are shown sampling from a bivariate Gaussian distribution with high correlation. Gibbs sampling undertakes a random walk, and in the 40 iter-ations shown (each consisting of an update of both variables) succeeds in moving only a small way along the long axis of the distribution. In the same number of iterations, Adler's Gaussian overrelaxation method with $\alpha = -0.98$ covers a greater portion of the distribution, since it tends to move consistently in one direction (subject to some random variation, and to "reflection" from the end of the distribution).

The manner in which overrelaxation avoids doing a random walk when sampling from this distribution is illustrated in the close-up view in Figure 1, which shows the changes in state after each variable is updated. When each of these updates is overrelaxed — tending to move to the other side of the conditional mean — the combined effect is to move in a consistent direction. The effect can be visualized most easily when $\alpha = -1$, in which case the state stays on a single elliptic contour of the probability density. Successive updates move the state along this contour until the end is reached, at which point the motion reverses. When $\alpha$ is close to, but not quite, $-1$, the small amount of randomness introduced will let the state move to different contours, and will also cause occasional reversals in direction of motion.

When $\alpha$ is chosen well, this randomization will occur on about the same time scale as is required for the state to move from one end of the distribution to the other. As the correlation of the bivariate Gaussian approaches $\pm 1$, the optimal value of $\alpha$ approaches $-1$, and the benefit from using this optimal $\alpha$ rather than $\alpha = 0$ (Gibbs sampling) becomes arbitrarily large. This comes about because the number, $n$, of typical steps required to move from one end of the distribution to the other is proportional to the square root of the ratio of eigenvalues of the correlation matrix, which goes to infinity as the correlation goes to $\pm 1$. The gain from moving to a nearly independent point in $n$ step rather than the $n^2$ steps needed with a random walk therefore also goes to infinity.

### 2.3 The benefit from overrelaxation

Figure 2 shows the benefit of overrelaxation in sampling from the bivariate Gaussian with $\rho = 0.998$ in terms of reduced autocorrelations for two functions of the state, $x_1$, and $x_1^2$. Here, a value of $\alpha = -0.89$ was used, which is close to optimal in terms of speed of convergence for this distribution. (The value of $\alpha = -0.98$ in Figure 1 was chosen to make the suppression of random walks visually clearer, but it is in fact somewhat too extreme for this value of $\rho$.)

The asymptotic efficiency of a Markov chain sampling method in estimating the expectation of a function of state is given by its "autocorrelation time" — the sum of the autocorrelations for that function of state at all lags (positive and negative). I obtained numerical estimates of the autocorrelation times for the quantities plotted in Figure 2 (using a series of 10,000 points, with truncation at the lags past which the estimated autocorrelations appeared to be approximately zero). These estimates show that the efficiency of estimation of $E[x_1]$ is a factor of about 22 better when using overrelaxation with $\alpha = -0.89$ than when using Gibbs sampling. For estimation of $E[x_1^2]$, the benefit from overrelaxation is a factor of about 16. In comparison with Gibbs sampling, using overrelaxation will reduce by these factors the variance of an estimate that is based on a run of given length, or alternatively, it will reduce by the same factors the length of run

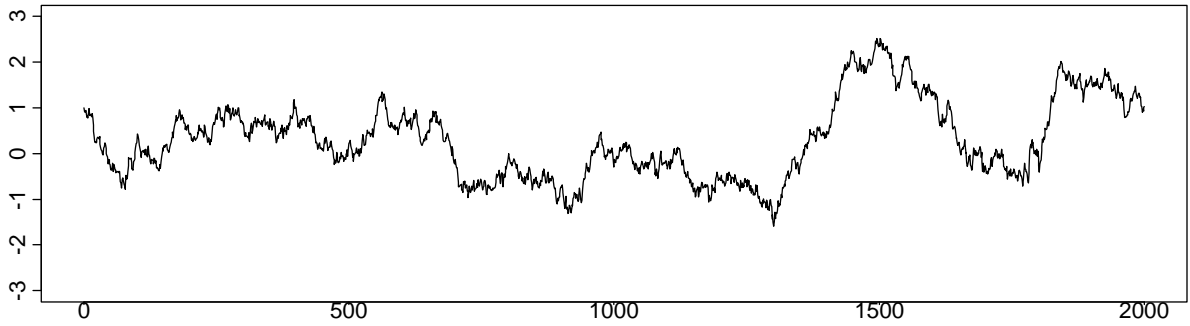

*Plot of $x_1$ during Gibbs sampling run*

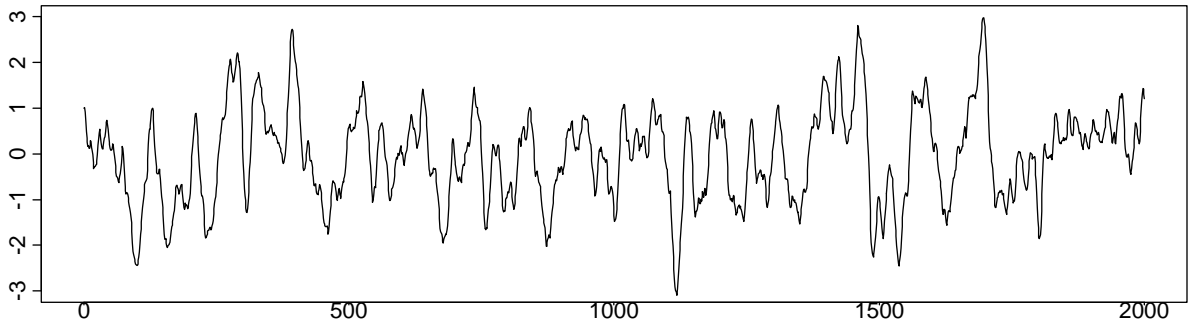

*Plot of $x_1$ during overrelaxed run with $\alpha = -0.89$*

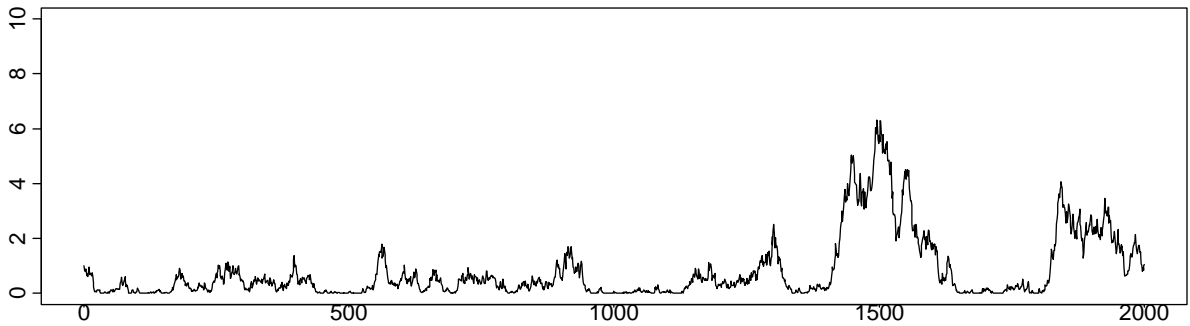

*Plot of $x_1^2$ during Gibbs sampling run*

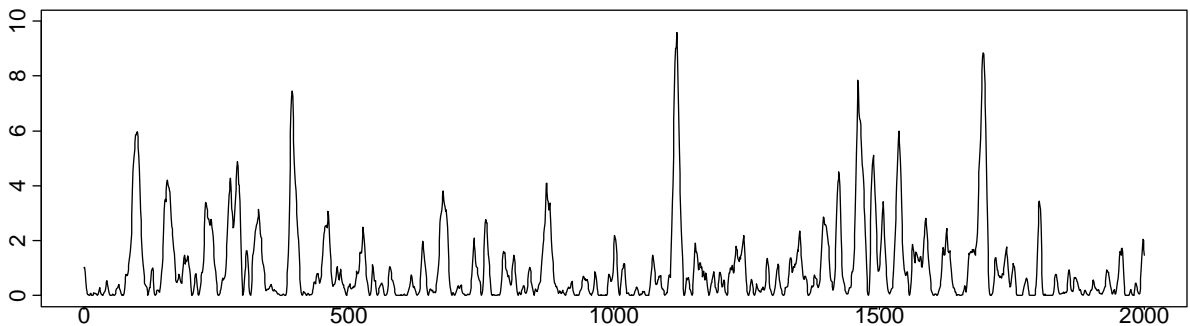

*Plot of $x_1^2$ during overrelaxed run with $\alpha = -0.89$*

Figure 2: Sampling from a bivariate Gaussian with $\rho = 0.998$ using Gibbs sampling and Adler's overrelaxation method with $\alpha = -0.89$. The plots show the values of the first coordinate and of its square during 2000 iterations of the samplers (each iteration consisting of one update for each coordinate).

that is required to reduce the variance to some desired level.

Overrelaxation is not always beneficial, however. Some research has been done into when overrelaxation produces an improvement, but the results obtained so far do not provide a complete answer, and in some cases, appear to have been mis-interpreted. Work in the physics literature has concentrated on systems of physical interest, and has primarily been concerned with scaling behaviour in the vicinity of a critical point. Two recent papers have addressed the question in the context of more general statistical applications.

Barone and Frigessi (1990) look at overrelaxation applied to multivariate Gaussian distributions, finding the rate of convergence in a number of interesting cases. In interpreting these results, however, one should keep in mind that if a method converges geometrically with rate $\rho$, the computation time required to reach some given level of accuracy is inversely proportional to $-\log(\rho)$. Hence, if two methods converge with rates $\rho_1$ and $\rho_2$, the relative advantage of method 2 is not $\rho_1/\rho_2$, but rather $\log(\rho_2)/\log(\rho_1)$, which for rates near one is approximately $(1-\rho_2)/(1-\rho_1)$. Seen in this light, the results of Barone and Frigessi confirm what was illustrated above — that overrelaxation can for some distributions be arbitrarily faster than Gibbs sampling. This is not always true, however; indeed, their results show that for some distributions with negative correlations, it can be better to underrelax (ie, to apply equation (1) with $0 < \alpha < 1$).

Green and Han (1992) look at the performance of overrelaxation as judged by the asymptotic variance of a Monte Carlo estimate for the expectation of a linear function of state. They show that this asymptotic variance goes to zero in the limit as $\alpha$ in equation (1) goes to $-1$. Recognizing that values for $\alpha$ very near $-1$ are not good from the point of view of convergence to equilibrium, since the state then remains for a long time on one contour of the joint probability density, they suggest that different chains be used during an initial period when equilibrium is being reached and during the subsequent generation of states for use in estimation.

In practice, however, we are usually interested in a non-linear function of state, and we require only a modest degree of accuracy. The results of Green and Han on asymptotic variance may therefore be of little relevance. In particular, for problems where Markov chain sampling is necessary, it is generally unrealistic to hope to find "antithetic" methods, in which negative autocorrelations result in estimation efficiencies greater than would be obtained with a sample of independent states. Fortunately, despite the locally-antithetic character of equation (1), the benefits of overrelaxation are not dependent on negative autocorrelations carrying over to the functions of interest, but can come rather from the faster decay of autocorrelations to zero, as the chain moves more rapidly to a nearly independent state.

As remarked above, the benefits of overrelaxation are not universal, even within the class of multivariate Gaussian distributions (despite the results on asymptotic variance). More research into this matter is needed. In this work, however, I take it as given that in many contexts overrelaxation is beneficial — as is typically true when correlations between components of the state are positive, for example — and seek to extend these benefits to distributions where the conditional distributions are non-Gaussian.

# 3   Previous proposals for more general overrelaxation methods

Adler's overrelaxation method can be applied only when all the full conditional distributions are Gaussian. Although applications of this nature do exist, in both statistical physics and statistical inference, most problems to which Markov chain Monte Carlo methods are currently applied do not satisfy this constraint. A number of proposals have been made for more general overrelaxation methods, which I will review here before presenting the "ordered overrelaxation" method in the next section.

Brown and Woch (1987) make a rather direct proposal: To perform an overrelaxed update of a variable whose conditional distribution is not Gaussian, transform to a new parameterization of this variable in which the conditional distribution is Gaussian, do the update by Adler's method, and then transform back. This may sometimes be an effective strategy, but for many problems the required computations will be costly or infeasible.

A second proposal by Brown and Woch (1987), also made by Creutz (1987), is based on the Metropolis algorithm. To update component $i$, we first find a point, $x_i^*$, which is near the centre of the conditional conditional distribution, $\pi(x_i \,|\, \{x_j\}_{j \neq i})$. We might, for example, choose $x_i^*$ to be an approximation to the mode, though other choices are also valid, as long as they do not depend on the current $x_i$. We then take $x_i' = x_i^* - (x_i - x_i^*)$ as a candidate for the next state, which, in the usual Metropolis fashion, we accept with probability $\min[1,\, \pi(x_i' \,|\, \{x_j\}_{j \neq i}) \,/\, \pi(x_i \,|\, \{x_j\}_{j \neq i})]$. If $x_i'$ is not accepted, the new state is the same as the old state.

If the conditional distribution is Gaussian, and $x_i^*$ is chosen to be the exact mode, the state proposed with this method will always be accepted, since the Gaussian distribution is symmetrical. The result is then identical to Adler's method with $\alpha = -1$. Such a method can be combined with other updates to produce an ergodic chain. Alternatively, ergodicity can be ensured by adding some amount of random noise to the proposed states.

Green and Han (1992) propose a somewhat similar, but more general, method. To update component $i$, they find a Gaussian approximation to the conditional distribution, $\pi(x_i \,|\, \{x_j\}_{j \neq i})$, that does not depend on the current $x_i$. They then find a candidate state $x_i'$ by overrelaxing from the current state according to equation (1), using the $\mu_i$ and $\sigma_i$ that characterize this Gaussian approximation, along with some judiciously chosen $\alpha$. This candidate state is then accepted or rejected using Hastings' (1970) generalization of the Metropolis algorithm, which allows for non-symmetric proposal distributions.

Fodor and Jansen (1994) propose a method that is applicable when the conditional distribution is unimodal, in which the candidate state is the point on the other side of the mode whose probability density is the same as that of the current state. This candidate state is accepted or rejected based on the derivative of the mapping from current state to candidate state. Ergodicity may again be ensured by mixing in other transitions, such as standard Metropolis updates.

The proposed generalizations in which detailed balance is achieved using accept-reject decisions all suffer from a potentially serious flaw: The rejection rate is determined by characteristics of the conditional distributions; if it is too high, there is no obvious way of reducing it. Moreover, even a quite small rejection rate may be too high. This point seems not to have been appreciated in the literature, but should be apparent from the discussion in Section 2. When sampling from a bivariate Gaussian, it is easy to see that when an overrelaxed update of one of the two variables is rejected, the effect is to reverse the direction of motion along the long axis of the distribution. Effective suppression of random walks therefore requires that the interval between such rejections be at least comparable to the time required for the method to move the length of the distribution, which can be arbitrarily long, depending on the degree of correlation of the variables.

## 4 Overrelaxation based on order statistics

In this section, I present a new form of overrelaxation, which can be applied (in theory) to any distribution over states with real-valued components, and in which changes are never rejected, thereby preserving the potential for the method to suppress random walks even in distributions with arbitrarily strong dependencies.

### 4.1 The ordered overrelaxation method

As before, we aim to sample from a distribution over $x = (x_1, \ldots, x_N)$ with density $\pi(x)$, and we will proceed by updating the values of the components, $x_i$, repeatedly in turn, based on their full conditional distributions, whose densities are $\pi(x_i \,|\, \{x_j\}_{j \neq i})$.

In the new method, the old value, $x_i$, for component $i$ is replaced by a new value, $x'_i$, obtained as follows:

1) Generate $K$ random values, independently, from the conditional distribution $\pi(x_i \,|\, \{x_j\}_{j \neq i})$.

2) Arrange these $K$ values plus the old value, $x_i$, in non-decreasing order, labeling them as follows:
$$x_i^{(0)} \leq x_i^{(1)} \leq \cdots \leq x_i^{(r)} = x_i \leq \cdots \leq x_i^{(K)} \qquad (2)$$
with $r$ being the index in this ordering of the old value. (If several of the $K$ generated values are equal to the old $x_i$, the tie is broken at random.)

3) Let the new value for component $i$ be $x'_i = x_i^{(K-r)}$.

Here, $K$ is a parameter of the method, which plays a role analogous to that of $\alpha$ in Adler's Gaussian overrelaxation method. When $K$ is one, the method is equivalent to Gibbs sampling; the behaviour as $K \to \infty$ is analogous to Gaussian overrelaxation with $\alpha = -1$.

As presented above, each step of this "ordered overrelaxation" method would appear to require computation time proportional to $K$. As discussed below, the method will provide a practical improvement in sampling efficiency only if an equivalent effect can be obtained using much less time. Strategies for accomplishing this are discussed in Section 5. First, however, I will show that the method is valid — that the update described above leaves the distribution $\pi(x)$ invariant — and that its behaviour is similar to that of Adler's method for Gaussian distributions.

## 4.2 Validity of ordered overrelaxation

To show that ordered overrelaxation leaves $\pi(x)$ invariant, it suffices to show that each update for a component, $i$, satisfies "detailed balance" — ie, that the probability density for such an update replacing $x_i$ by $x'_i$ is the same as the probability density for $x'_i$ being replaced by $x_i$, assuming that the starting state is distributed according to $\pi(x)$. It is well known that the detailed balance condition (also known as "reversibility") implies invariance of $\pi(x)$, and that invariance for each component update implies invariance for transitions in which each component is updated in turn. (Note, however, that the resulting sequential update procedure, considered as a whole, need not satisfy detailed balance; indeed, if random walks are to be suppressed as we wish, it must not.)

To see that detailed balance holds, consider the probability density that component $i$ has a given value, $x_i$, to start, that $x_i$ is in the end replaced by some given different value, $x'_i$, and that along the way, a particular set of $K-1$ other values (along with $x'_i$) are generated in step (1) of the update procedure. Assuming there are no tied values, this probability density is

$$\pi(x_i \,|\, \{x_j\}_{j \neq i}) \cdot K!\, \pi(x'_i \,|\, \{x_j\}_{j \neq i}) \prod_{r \neq t \neq s} \pi(x_i^{(t)} \,|\, \{x_j\}_{j \neq i}) \cdot I[s = K-r] \qquad (3)$$

where $r$ is the index of the old value, $x_i$, in the ordering found in step (2), and $s$ is the index of the new value, $x'_i$. The final factor is zero or one, depending on whether the transition in question would actually occur with the particular set of $K-1$ other values being considered. The probability density for the reverse transition, from $x'_i$ to $x_i$, with the same set of $K-1$ other values being involved, is readily seen to the identical to the above. Integrating over all possible sets of other values, we conclude that the probability density for a transition from $x_i$ to $x'_i$, involving any set of other values, is the same as the probability density for the reverse transition from $x'_i$ to $x_i$. Allowing for the possibility of ties yields the same result, after a more detailed accounting.

## 4.3 Behaviour of ordered overrelaxation

In analysing ordered overrelaxation, it can be helpful to view it from a perspective in which the overrelaxation is done with respect to a uniform distribution. Let $F(x)$ be the cumulative distribution function for the conditional distribution $\pi(x_i \,|\, \{x_j\}_{j \neq i})$ (here assumed to be continuous), and let $F^{-1}(x)$ be the inverse of $F(x)$. Ordered overrelaxation for $x_i$ is equivalent to the following procedure: First transform the current value to $u_i = F(x_i)$, then perform ordered overrelaxation for $u_i$, whose distribution is uniform over $[0, 1]$, yielding a new state $u'_i$, and finally transform back to $x'_i = F^{-1}(u'_i)$.

Overrelaxation for a uniform distribution, starting from $u$, may be analysed as follows. When $K$ independent uniform variates are generated in step (1) of the procedure, the number of them that are less than $u$ will be binomially distributed with mean $Ku$ and variance $Ku(1-u)$. This number is the index, $r$, of $u = u^{(r)}$ found in step (2) of the procedure. Conditional on a value for $r$, which let us suppose is greater than $K/2$, the distribution of the new state, $u' = u^{(K-r)}$, will be that of the $K-r+1$ order statistic of a sample of size $r$ from a uniform distribution over $[0, u]$. As is well known (eg, David 1970, p. 11), the $k$'th order statistic of a sample of size $n$ from a uniform distribution over $[0, 1]$ has a beta($k$, $n-k+1$) distribution, with density proportional to $u^{k-1}(1-u)^{n-k}$, mean $k/(n+1)$, and variance $k(n-k+1)/(n+2)(n+1)^2$. Applying this result, $u'$ for a given $r > K/2$ will have a rescaled beta($K-r+1$, $2r-K$) distribution, with mean $\mu(r) = u(K-r+1)/(r+1)$ and variance $\sigma^2(r) = u^2(K-r+1)(2r-K)/(r+2)(r+1)^2$.

When $K$ is large, we can get a rough idea of the behaviour of overrelaxation for a uniform distribution by considering the case where $u$ (and hence likely $r/K$) is significantly greater than $1/2$. Behaviour when $u$ is significantly less than $1/2$ will of course be symmetrical, and we expect behaviour to smoothly interpolate between these regimes when $u$ is within about $1/\sqrt{K}$ of $1/2$ (for which $r/K$ might be either greater or less than $1/2$)

When $u \gg 1/2$, we can use the Taylor expansion

$$\mu(Ku+\delta) \;=\; \frac{Ku - Ku^2 + u}{Ku+1} \;-\; \frac{Ku+2u}{(Ku+1)^2}\delta \;+\; \frac{Ku+2u}{(Ku+1)^3}\delta^2 \;+\; \cdots \qquad (4)$$

to conclude that for large $K$, the expected value of $u'$, averaging over possible values for $r = Ku + \delta$, with $\delta$ having mean zero and variance $Ku(1-u)$, is approximately

$$\frac{Ku - Ku^2 + u}{Ku+1} \;+\; Ku(1-u)\frac{Ku+2u}{(Ku+1)^3} \;\approx\; (1-u) \;+\; 1/K \qquad (5)$$

For $u \ll 1/2$, the bias will of course be opposite, with the expected value of $u'$ being about $(1-u) - 1/K$, and for $u \approx 1/2$, the expected value of $u'$ will be approximately $u$.

The variance of $u'$ will be, to order $1/K$, approximately $\sigma^2(Ku) + Ku(1-u)[\mu'(Ku)]^2$; that is, for $u \gg 1/2$:

$$\frac{(Ku^2 - Ku^3 + u^2)(2Ku - K)}{(Ku+2)(Ku+1)^2} \;+\; Ku(1-u)\frac{(Ku+2u)^2}{(Ku+1)^4} \;\approx\; 2(1-u)/K \qquad (6)$$

By symmetry, the variance of $u'$ when $u \ll 1/2$ will be approximately $2u/K$. (Incidentally, the fact that $u'$ has greater variance when $u$ is near $1/2$ than when $u$ is near 0 or 1 explains how it is possible for the method to leave the uniform distribution invariant even though $u'$ is biased to be closer to $1/2$ than $u$ is.)

The joint distribution for $u$ and $u'$ is illustrated on the left in Figure 3. The right of the figure shows how this translates to the joint distribution for the old and new state when ordered overrelaxation is applied to a Gaussian distribution.

### 4.4 Comparisons with Gaussian overrelaxation and Gibbs sampling

For Gaussian overrelaxation by Adler's method, the joint distribution of the old and new state is Gaussian. As seen in Figure 3, this is clearly not the case for ordered overrelaxation. One notable difference is the way the tails of the joint distribution flare out with ordered overrelaxation, a reflection of the fact that if the old state is very far out in the tail, the new state will likely be much closer in. This effect is perhaps an advantage of the ordered overrelaxation method, as one might therefore expect convergence from a bad starting point to be faster with ordered overrelaxation than with Adler's method. (This is certainly true in the trivial case where the state consists of a single variable; further analysis is needed to establish whether it true in interesting cases.)

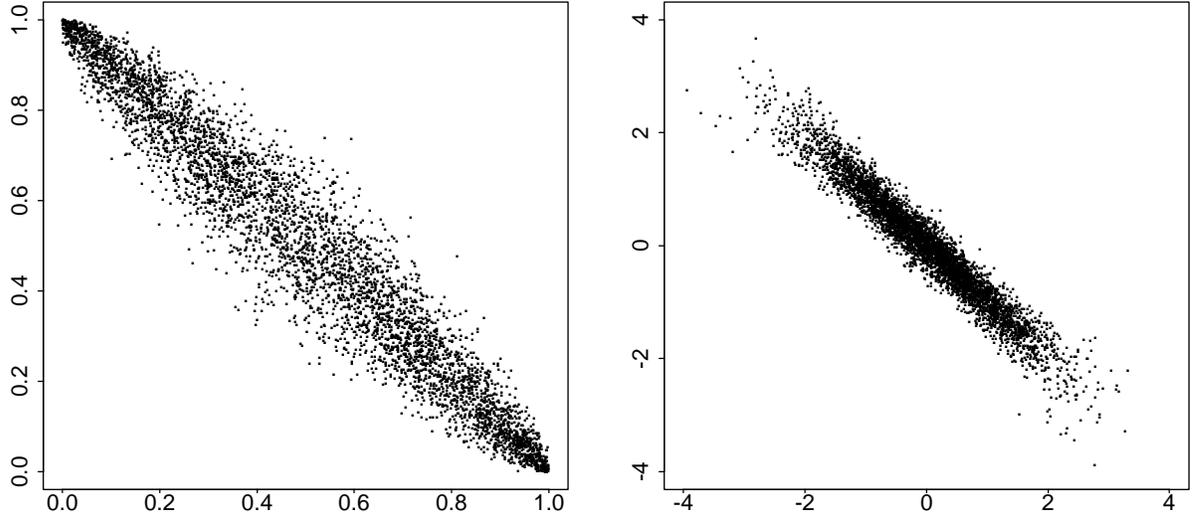

Figure 3: Points representing 5000 ordered overrelaxation updates. The plot on the left shows ordered overrelaxation for a uniform distribution. The horizontal axis gives the starting point, drawn uniformly from [0, 1]; the vertical axis, the point found by ordered overrelaxation with $K = 100$ from that starting point. The plot on the right shows ordered overrelaxation for a Gaussian distribution. The points correspond to those on the left, but transformed by the inverse Gaussian cumulative distribution function.

Although there is no exact equivalence between Adler's Gaussian overrelaxation method and ordered overrelaxation, it is of some interest to find a value of $K$ for which ordered overrelaxation applied to a Gaussian distribution corresponds roughly to Adler's method with a given $\alpha < 0$. Specifically, we can try to equate the mean and variance of the new state, $x'$, that results from an overrelaxed update of an old state, $x$, when $x$ is one standard deviation away from its mean. Supposing without loss of generality that the mean is zero and the variance is one, we see from equations (5) and (6) that when $x = 1$, the expected value of $x'$ using ordered overrelaxation is $\Phi^{-1}(\Phi(-1) + 1/K) \approx -1 + 1/K\phi(-1) \approx -1 + 4.13/K$ and the variance of $x'$ is $2\Phi(-1)/K\phi(-1)^2 \approx 5.42/K$, where $\Phi(x)$ is the Gaussian cumulative distribution function, and $\phi(x)$ the Gaussian density function. Since the corresponding values for Adler's method are a mean of $\alpha$ and a variance of $1 - \alpha^2$, we can get a rough correspondence by setting $K \approx 3.5/(1 + \alpha)$.

For the example of Figure 2, showing overrelaxation by Adler's method with $\alpha = -0.89$, applied to a bivariate Gaussian with correlation 0.998, ordered overrelaxation should be roughly equivalent when $K = 32$. Figure 4 shows visually that this is indeed the case. Numerical estimates of autocorrelation times indicate that ordered overrelaxation with $K = 32$ is about a factor of 22 more efficient, in terms of the number of iterations required for a given level of accuracy, than is Gibbs sampling, when used to estimate $E[x_1]$. When used to estimate $E[x_1^2]$, ordered overrelaxation is about a factor of 14 more efficient. Measured by numbers of iterations, these efficiency advantages are virtually identical to those reported in Section 2.3 for Adler's method.

Of course, if it were implemented in the most obvious way, with $K$ random variates being explicitly generated in step (1) of the procedure, ordered overrelaxation with $K = 32$ would required a factor of about 32 more computation time per iteration than would

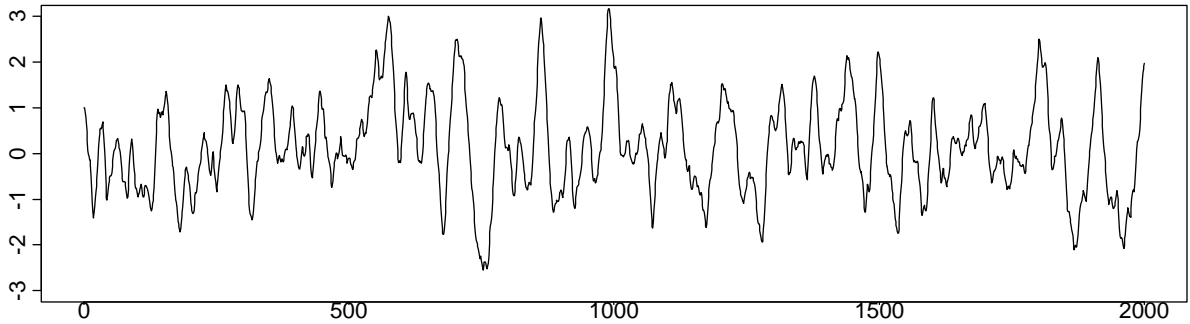

Plot of $x_1$ during ordered overrelaxation run with $K = 32$

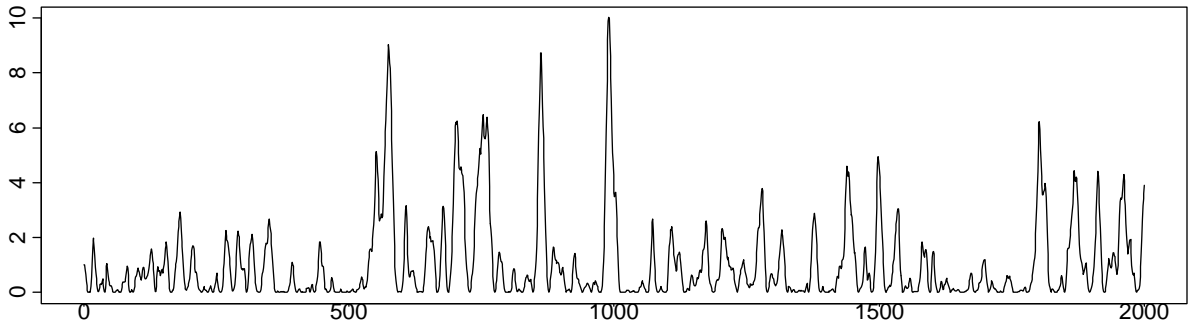

Plot of $x_1^2$ during ordered overrelaxation run with $K = 32$

Figure 4: Sampling from a bivariate Gaussian with $\rho = 0.998$ using ordered overrelaxation with $K = 32$. Compare with the results using Gibbs sampling and Adler's method shown in Figure 2.

either Adler's overrelaxation method or Gibbs sampling. Adler's method would clearly be preferred in comparison to such an implementation of ordered overrelaxation. Interestingly, however, even with such a naive implementation, the computational efficiency of ordered overrelaxation is comparable to that of Gibbs sampling — the factor of about 32 slowdown per iteration being nearly cancelled by the factor of about 22 improvement from the elimination of random walks. This near equality of costs holds for smaller values of $K$ as well — the improvement in efficiency (in terms of iterations) using ordered overrelaxation with $K = 16$ is about a factor of 12 for $E[x_1]$ and 11 for $E[x_1^2]$, and with $K = 8$, the improvement is about a factor of 8 for $E[x_1]$ and 7 for $E[x_1^2]$.

We therefore see that any implementation of ordered overrelaxation whose computational cost is substantially less than that of the naive approach of explicitly generating $K$ random variates will yield a method whose computational efficiency is greater than that of Gibbs sampling, when used with any value for $K$ up to that which is optimal in terms of the number of iterations required for a given level of accuracy. Or rather, we see this for the case of a bivariate Gaussian distribution, and we may hope that it is true for many other distributions of interest as well, including those whose conditional distributions are non-Gaussian, for which Adler's overrelaxation method is not applicable.

# 5 Strategies for implementing ordered overrelaxation

In this section, I describe several approaches to implementing ordered overrelaxation, which are each applicable to some interesting class of distributions, and are more efficient than the obvious method of explicitly generating $K$ random variates. In some cases, there is a bound on the time required for an overrelaxed update that is independent of $K$; in others the reduction in time is less dramatic (perhaps only a constant factor). As was seen in Section 4.4, any substantial reduction in time compared to the naive implementation will potentially provide an improvement over Gibbs sampling.

There will of course be some distributions for which none of these implementations is feasible; this will certainly be the case when Gibbs sampling itself is not feasible. Such distributions include, for example, the complex posterior distributions that arise with neural network models (Neal 1995). Hybrid Monte Carlo will likely remain the most efficient sampling method for such problems.

## 5.1 Implementation using the cumulative distribution function

The most direct method for implementing ordered overrelaxation in bounded time (independently of $K$) is to transform the problem to one of performing overrelaxation for a uniform distribution on $[0, 1]$, as was done in the analysis of Section 4.3. This approach requires that we be able to efficiently compute the cumulative distribution function and its inverse for each of the conditional distributions for which overrelaxation is to be done. This requirement is somewhat restrictive, but reasonably fast methods for computing these functions are known for many standard distributions (Kennedy and Gentle 1980).

This implementation of ordered overrelaxation produces exactly the same effect as would a direct implementation of the steps in Section 4.1. As there, our aim is to update the current value, $x_i$, for component $i$, replacing it with a new value, $x_i'$. The conditional distribution for component $i$, $\pi(x_i \mid \{x_j\}_{j \neq i})$, is here assumed to be continuous, with cumulative distribution distribution function $F(x)$, whose inverse is $F^{-1}(x)$. We proceed as follows:

1) Compute $u = F(x_i)$, which will lie in $[0, 1]$.

2) Randomly draw an integer $r$ from the binomial$(K, u)$ distribution. This $r$ has the same distribution as the $r$ in the direct procedure of Section 4.1.

3) If $r > K - r$, randomly generate $v$ from the beta$(K - r + 1, 2r - K)$ distribution, and let $u' = uv$.

   If $r < K - r$, randomly generate $v$ from the beta$(r + 1, K - 2r)$ distribution, and let $u' = 1 - (1 - u)v$.

   If $r = K - r$, let $u' = u$.

   Note that $u'$ is the result of overrelaxing $u$ with respect to the uniform distribution on $[0, 1]$.

4) Let the new value for component $i$ be $x_i' = F^{-1}(u')$.

Step (3) is based on the fact (David 1970, p. 11) that the $k$'th order statistic in a sample of size $n$ from a uniform distribution on $[0, 1]$ has a beta($k$, $n - k + 1$) distribution. Efficient methods for generating random variates from beta and binomial distributions in bounded expected time are known (Devroye 1986, Sections IX.4 and X.4).

When feasible, this implementation allows ordered overrelaxation to be performed in time independent of $K$, though this time will exceed that required for a simple Gibbs sampling update. The implementation is similar in spirit and in likely computation time to the suggestion by Brown and Woch (1987) to perform a transformation that makes the conditional distribution Gaussian (see Section 3). The ordered overrelaxation framework admits other possible implementations, however, which may reduce the required computation time, or allow its application to distributions for which the cumulative distribution function or its inverse cannot be computed.

## 5.2 Implementations based on economies of scale

In some cases, an ordered overrelaxation update will quite naturally take less than $K$ times as long as a Gibbs sampling update (and therefore be potentially advantageous). In the direct procedure of Section 4.1, the conditional distribution for $x_i$, from which $K$ random variates are drawn in step (1), depends in general on the values of the other components, perhaps in a complex way. Since these other components are themselves being updated, this conditional distribution must be re-computed for each update of $x_i$. This need be done only once for an ordered overrelaxation update, however, even though $K$ values will then be generated from the conditional distribution that is found. If the dominant contribution to the total computation time comes from this dependence on the values of the other components, rather than from the random variate generation itself, an ordered overrelaxation update could take much less than $K$ times as long as a Gibbs sampling update.

Other situations can also lead to "economies of scale", in which generating $K$ values from the same distribution takes less than $K$ times as long as generating one value. This will occur whenever values are generated from some distribution in a parametric family using a method with some "setup cost" that is incurred whenever the parameters change (due to dependence on other components of state). The adaptive rejection sampling method of Gilks and Wild (1992) is another important example, as it is widely used to implement Gibbs sampling. In this scheme, a value is randomly drawn from a log-concave density using a succession of approximations to the density function. When more than one value is drawn from the same density, the approximations are continually refined, with the result that later values take much less time to generate than earlier values.

Further time savings can be obtained by noting that the exact numerical values of most of the $K$ values generated are not needed. All that is required is that the number, $r$, of these values that are less than the current $x_i$ be somehow determined, and that the single value $x_i^{(K-r)} = x_i'$ be found. In particular, the adaptive rejection sampling method can be modified in such a way that large groups of values are "generated" only to the extent that they are localized to regions where their exact values can be seen to be irrelevant. The cost of ordered overrelaxation can then be much less than $K$ times the cost of a Gibbs sampling update. The details of this procedure and its analysis are somewhat complex, however; I plan to present them in a future paper.

## 6 Demonstration: Inference for a hierarchical Bayesian model

In this section, I demonstrate the advantages of ordered overrelaxation over Gibbs sampling when both are applied to Bayesian inference for a simple hierarchical model. In this problem, the conditional distributions are non-Gaussian, so Adler's method cannot be applied. The implementation of ordered overrelaxation used is that based on the cumulative distribution function, described in Section 5.1.

For this demonstration, I used one of the models Gelfand and Smith (1990) use to illustrate Gibbs sampling. The data consist of $p$ counts, $s_1, \ldots, s_p$. Conditional on a set of unknown underlying parameters, $\lambda_1, \ldots, \lambda_p$, these counts are assumed to have independent Poisson distributions, with means of $\lambda_i t_i$, where the $t_i$ are known quantities associated with the counts $s_i$. For example, $s_i$ might be the number of failures of a device that has a failure rate of $\lambda_i$ and that has been observed for a period of time $t_i$.

At the next level, a common hyperparameter $\beta$ is introduced. Conditional on a value for $\beta$, the $\lambda_i$ are assumed to be independently generated from a gamma distribution with a known shape parameter, $\alpha$, and the scale factor $\beta$. The hyperparameter $\beta$ itself is assumed to have an inverse gamma distribution with a known shape parameter, $\gamma$, and a known scale factor, $\delta$.

The problem is to sample from the conditional distribution for $\beta$ and the $\lambda_i$ given the observed $s_1, \ldots, s_p$. The joint density of all unknowns is given by the following proportionality:

$$P(\beta, \lambda_1, \ldots, \lambda_p \mid s_1, \ldots, s_p) \;\propto\; P(\beta)\, P(\lambda_1, \ldots, \lambda_p \mid \beta)\, P(s_1, \ldots, s_p \mid \lambda_1, \ldots, \lambda_p) \quad (7)$$

$$\propto\; \beta^{-\gamma-1} e^{-\delta/\beta} \cdot \prod_{i=1}^{p} \beta^{-\alpha} \lambda_i^{\alpha-1} e^{-\lambda_i/\beta} \cdot \prod_{i=1}^{p} \lambda_i^{s_i} e^{-\lambda_i t_i} \quad (8)$$

The conditional distribution for $\beta$ given the other variables is thus inverse gamma:

$$P(\beta \mid \lambda_1, \ldots, \lambda_p, s_1, \ldots, s_p) \;\propto\; \beta^{-p\alpha - \gamma - 1} e^{-(\delta + \Sigma_i \lambda_i)/\beta} \quad (9)$$

However, I found it more convenient to work in terms of $\tau = 1/\beta$, whose conditional density is gamma:

$$P(\tau \mid \lambda_1, \ldots, \lambda_p, s_1, \ldots, s_p) \;\propto\; \tau^{p\alpha + \gamma - 1} e^{-\tau(\delta + \Sigma_i \lambda_i)} \quad (10)$$

The conditional distributions for the $\lambda_i$ are also gamma:

$$P(\lambda_i \mid \{\lambda_j\}_{j \neq i}, \tau, s_1, \ldots, s_p) \;\propto\; \lambda_i^{s_i + \alpha - 1} e^{-\lambda_i(t_i + \tau)} \quad (11)$$

In each full iteration of Gibbs sampling or of ordered overrelaxation, these conditional distributions are used to update first the $\lambda_i$ and then $\tau$.

Gelfand and Smith (1990, Section 4.2) apply this model to a small data set concerning failures in ten pump systems, and find that Gibbs sampling essentially converges within ten iterations. Such rapid convergence does not always occur with this model, however.

The $\lambda_i$ and $\tau$ are mutually dependent, to a degree that increases as $\alpha$ and $p$ increase. By adjusting $\alpha$ and $p$, one can arrange for Gibbs sampling to require arbitrarily many iterations to converge.

For the tests reported here, I set $p = 100$, $\alpha = 20$, $\delta = 1$, and $\gamma = 0.1$. The true value of $\tau$ was set to 5 (ie, $\beta = 0.2$). For each $i$ from 1 to $p$, $t_i$ was set to $i/p$, a value for $\lambda_i$ was randomly generated from the gamma distribution with parameters $\alpha$ and $\beta$, and finally a synthetic observation, $s_i$, was randomly generated from the Poisson distribution with mean $\lambda_i t_i$. A single such set of 100 observations was used for all the tests, during which the true values of $\tau$ and the $\lambda_i$ used to generate the data were of course ignored.

Figure 5 shows values of $\tau$ sampled from the posterior distribution by successive iterations of Gibbs sampling, and of ordered overrelaxation with $K = 5$, $K = 11$, and $K = 21$. Each of these methods was initialized with the $\lambda_i$ set to $s_i/t_i$ and $\tau$ set to $\alpha$ divided by the average of the initial $\lambda_i$. The ordered overrelaxation iterations took about 1.7 times as long as the Gibbs sampling iterations. (Although approximately in line with expectations, this timing figure should not be taken too seriously – since the methods were implemented in S-Plus, the times likely reflect interpretative overhead, rather than intrinsic computational difficulty.)

The figure clearly shows the reduction in autocorrelation for $\tau$ that can be achieved by using ordered overrelaxation rather than Gibbs sampling. Numerical estimates of the autocorrelations (with the first 50 points discarded) show that for Gibbs sampling, the autocorrelations do not approach zero until around lag 28, whereas for ordered overrelaxation with $K = 5$, the autocorrelation is near zero by lag 11, and for $K = 11$, by lag 4. For ordered overrelaxation with $K = 21$, substantial negative autocorrelations are seen, which would increase the efficiency of estimation for the expected value of $\tau$ itself, but could be disadvantageous when estimating the expectations of other functions of state. The value $K = 11$ seems close to optimal in terms of speed of convergence.

## 7 Discussion

The results in this paper show that ordered overrelaxation should be able to speed the convergence of Markov chain Monte Carlo in a wide range of circumstances. Unlike the original overrelaxation method of Adler (1981), it is applicable when the conditional distributions are not Gaussian, and it avoids the rejections that can undermine the performance of other generalized overrelaxation methods. Compared to the alternative of suppressing random walks using hybrid Monte Carlo (Duane, *et al.* 1987), overrelaxation has the advantage that it does not require the setting of a stepsize parameter, making it potentially easier to apply on a routine basis.

An implementation of ordered overrelaxation based on the cumulative distribution function was described in Section 5.1, and used for the demonstration in Section 6. This implementation can be used for many problems, but it is not as widely applicable as Gibbs sampling. Natural economies of scale will allow ordered overrelaxation to provide at least some benefit in many other contexts, without any special effort. By modifying adaptive rejection sampling (Gilks and Wild 1992) to rapidly perform ordered overrelax-

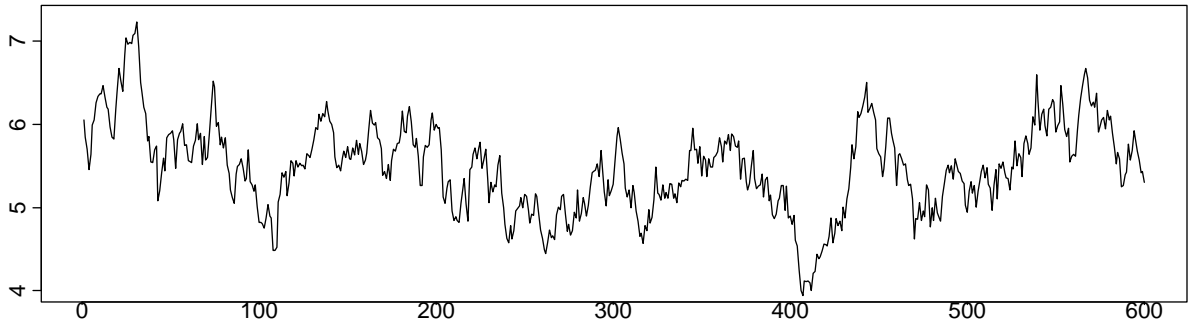
*Plot of $\tau$ during Gibbs sampling run*

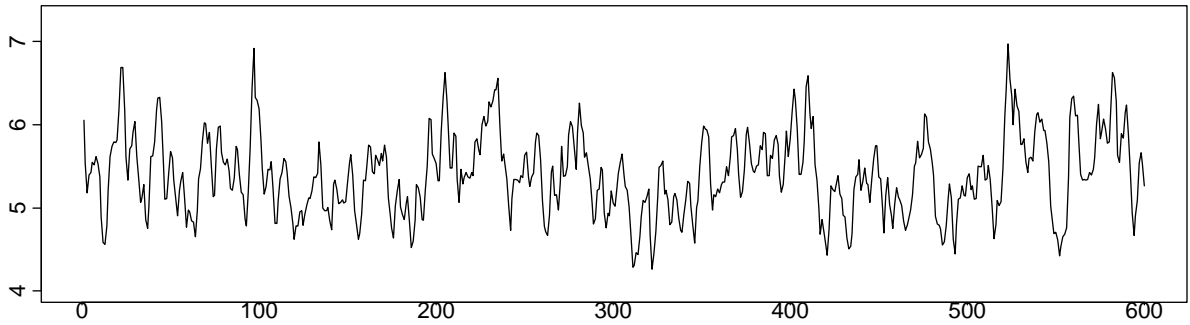
*Plot of $\tau$ during ordered overrelaxation run with $K = 5$*

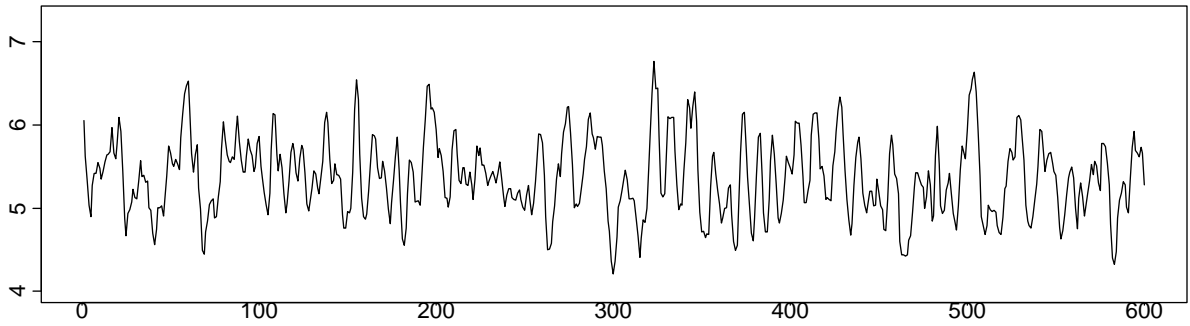
*Plot of $\tau$ during ordered overrelaxation run with $K = 11$*

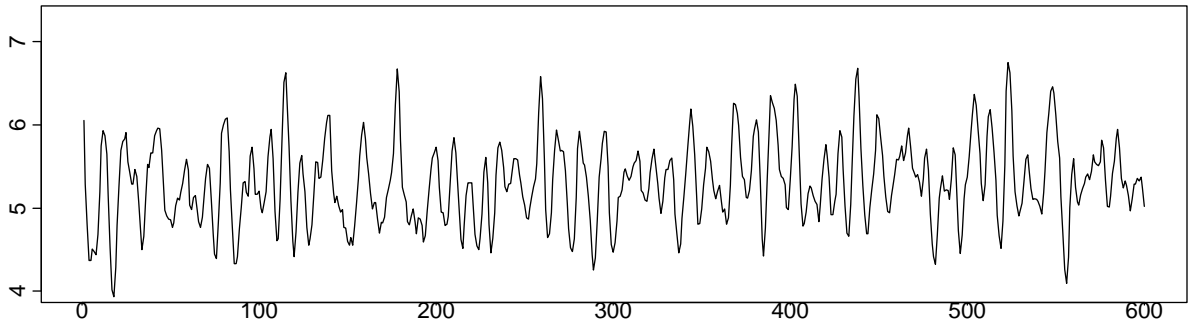
*Plot of $\tau$ during ordered overrelaxation run with $K = 21$*

Figure 5: Sampling from the posterior distribution for $\tau$ using Gibbs sampling and ordered overrelaxation with $K = 5$, $K = 11$, and $K = 21$. The plots show the progress of $\tau = 1/\beta$ during runs of 600 full iterations (in which the $\lambda_i$ and $\tau$ are each updated once).

ation, I believe that quite a wide range of problems will be able to benefit from ordered overrelaxation, which should often provide an order of magnitude or more speedup, with little effort on the part of the user.

To use overrelaxation, it is necessary for the user to set a time-constant parameter — $\alpha$ for Adler's method, $K$ for ordered overrelaxation — which, roughly speaking, controls the number of iterations for which random walks are suppressed. Ideally, this parameter should be set so that random walks are suppressed over the time scale required for the whole distribution to be traversed, but no longer. Short trial runs could be used to select a value for this parameter; finding a precisely optimal value is not crucial. In favourable cases, an efficient implementation of ordered overrelaxation used with any value of $K$ less than the optimal value will produce an advantage over Gibbs sampling of about a factor of $K$. Using a value of $K$ that is greater than the optimum will still produce an advantage over Gibbs sampling, up to around the point where $K$ is the square of the optimal value.

For routine use, a policy of simply setting $K$ to around 20 may be reasonable. For problems with a high degree of dependency, this may give around an order of magnitude improvement in performance over Gibbs sampling, with no effort by the user. For problems with little dependency between variables, for which this value of $K$ is too large, the result could be a slowdown compared with Gibbs sampling, but such problems are sufficiently easy anyway that this may cause little inconvenience. Of course, when convergence is very slow, or when many similar problems are to be solved, it will be well worthwhile to search for the optimal value of $K$.

There are problems for which overrelaxation (of whatever sort) is not advantageous, as can happen when variables are negatively correlated. Further research is needed to clarify when this occurs, and to determine how these situations are best handled. It can in fact be beneficial to underrelax in such a situation — eg, to use Adler's method with $\alpha > 0$ in equation (1). It is natural to ask whether there is an "ordered underrelaxation" method that could be used when the conditional distributions are non-Gaussian. I believe that there is. In the ordered overrelaxation method of Section 4.1, step (3) could be modified to randomly set $x'_i$ to either $x_i^{(K+1)}$ or $x_i^{(K-1)}$ (with the change being rejected if the chosen $K \pm 1$ is out of range). This is a valid update (preserving detailed balance), and should produce effects similar to those of Adler's method with $\alpha > 0$.

## Acknowledgements

I thank David MacKay for comments on the manuscript. This work was supported by the Natural Sciences and Engineering Research Council of Canada.